\documentstyle[12pt]{article}
\def\beq{\begin{eqnarray}}    
\def\eeq{\end{eqnarray}}      
\def\Tr{\,\mbox{Tr}\,}                  
\def\pa{\partial}                       
\def\sTr{\,\mbox{Str}\,}                
\def\sla{\!\!\!\slash}
\def\al{\alpha}
\def\be{\beta}

\def\ga{\gamma}
\def\de{\delta}
\def\vp{\varepsilon}
\def\ep{\epsilon}

\def\la{\lambda}
\def\na{\nabla}
\def\pa{\partial}

\def\si{\sigma}

\def\ph{\varphi}
\def\ta{\tau}
\def\th{\theta}

\def\Ga{\Gamma}
\def\De{\Delta}

\begin{document}
\newpage
\hfill CBPF-NF-061/99

\hfill DF/UFJF-1999/14

\hfill hep-th/9910168

\hfill JHEP 02(2000)003
\begin{center}

\vspace{1cm}
{\large\sc
On the Consistency of a Fermion-Torsion Effective Theory}

\vskip 8mm

\setcounter{page}1
\renewcommand{\thefootnote}{\arabic{footnote}}
\setcounter{footnote}0

{\bf G. de Berredo-Peixoto} $^{\mbox{a,}}$ \footnote{E-mail address:
peixoto@cbpf.br},
$\,\,$
{\bf J.A. Helayel-Neto} $^{\mbox{a,b,}}$ \footnote{E-mail address:
helayel@cbpf.br},
$\,\,$
{\bf I.L. Shapiro} $^{\mbox{c,d,}}$\footnote{E-mail address:
shapiro@ibitipoca.fisica.ufjf.br}

\vspace{7mm}

$^{\mbox{a}}$ {\it Centro Brasileiro de Pesquisas F\'{\i}sicas
(CBPF/CNPq), \\ Rio de Janeiro, Brazil} \vspace{3mm}

$^{\mbox{b}}$ {\it Universidade Cat\'olica de Petr\'opolis, 
\\ Petr\'opolis, Brazil}
\vspace{3mm}

{\it $^{\mbox{c}}\,\,$
Departamento de F\'\i sica - ICE,
Universidade Federal de Juiz de Fora 
\\ 36036-330, Juiz de Fora, MG, Brazil }
\vspace{3mm}

{\it $^{\mbox{d}}\,\,$
Tomsk State Pedagogical University, 634041, Tomsk, Russia }

\vspace{7mm}

\end{center}

\noindent
{\large\sl Abstract.}$\,\,\,\,$
We discuss the possibility to construct an effective quantum field
theory for an axial vector coupled to a Dirac spinor field. A massive
axial vector describes antisymmetric torsion. The consistency
conditions include unitarity and renormalizability in the
low-energy region. The investigation of the Ward identities 
and the one- and
two-loop divergences indicate serious problems arising in the
theory. The final conclusion is that torsion may exist as a string
excitation, but there are very severe restrictions for the existence of
a propagating torsion field, subject to the quantization
procedure, at low energies.

\newpage

\noindent 
{\large\bf Introduction} 
\vskip 2mm

With respect to the space-time symmetries, the Standard Model of
the Elementary Particle physics includes three types of fields:
spinors, vectors and scalars. The same concerns Grand Unified
Theories, which are indeed based on larger symmetry groups. The
effective interactions of QCD lead to the pion field, which is a
pseudo-scalar. One might, naturally, ask whether there may be other
fields or interactions which can be unobservable at low energies.
This question becomes particularly important in view of the fact 
that the
(super)string theories yield, in their low-energy spectrum,
some fields different from the ones mentioned above. Most of these
fields are not propagating (and, consequently, are not visible) at
available energies, because they have too huge masses (typically of
the Planck order). This concerns, at first, the higher-spin
excitations related to the massive string modes. Besides, in
addition to the 
usual fields, the massless excitations of the string
spectrum contain a skew-symmetric tensor,
which eventually produces, in the low-energy effective string
action, the 3-form associated to torsion. In known
string theories, this tensor shows up at first order in
$\al^{\prime}$ and has a mass of the Planck order. Therefore,
it doesn't propagate at low energies. However, it is interesting
to investigate the possibility that this field possesses 
an essentially
smaller (or zero) mass, so that torsion could propagate. This
implies the low theoretical bound for the torsion mass.

Here, we take the viewpoint
according to which any propagating field must
be quantized, so that the classical theory is nothing but an
approximation for the complete theory including quantum
corrections. Then, the appropriate framework for the investigation
of a propagating torsion is the effective
quantum field theory approach (see, for example, \cite{weinberg}).
From the modern point of view, most of the existing quantum field
theories should be regarded as effective ones, descending from 
some other
more fundamental theories. The classical action of the
effective theories may have the form of an infinite series
whose expansion is performed in the inverse of some
large massive parameter. At low energies, only the first
terms of the expansion are relevant, so that one
can consequently disregard high-derivative terms,
though some consistency conditions
should be indeed satisfied. In particular, the
theory must be unitary and renormalizable in the given low-energy
region. For the case of torsion, these consistency conditions
have been applied in \cite{betor}. It was shown that the
theory possesses an extra, softly broken, gauge symmetry and
that this symmetry fixes, in a unique way, the form of the
low-energy classical action. This action succeeds in the test
based on the calculation of the fermion determinant \cite{betor}
and led to a wide set of phenomenological consequences.

The purpose of the present paper is to proceed further with the study
of the possibility to construct a quantum field theory for a
fermion-torsion system. In \cite{betor},
the unique candidate to be torsion action was suggested and
some of its theoretical and phenomenological aspects were discussed.
It is well known that the theory of axial vector field
may have problems, and these problems are usually
accociated to the axial anomaly. However, in the case
of torsion embedding into the Standard Model, the anomaly
can not appear due to the algebraic reasons \cite{betor},
because all the vector ingredients of the SM have group
index which is absent for torsion. However, as the
example of the scalar-fermion-torsion shows, the absense
of anomaly does not guarantee consistency, and in particular 
the conflict between renormalizability and unitarity takes 
place.  Here we are going to investigate whether the
Ward-Takahashi identities and the one- and
two-loop divergences arising in the fermion-torsion system
are consistent with the requirements an
effective quantum field theory should fulfill.
This study is necessary for the final answer of whether the
space-time torsion can exist as an independent field,
propagating at low energies, which is
subject of quantization.

Our paper is organized as follows: in Section 2,
a brief review of the previous results is given
and the main purpose of the subsequent study is formulated.
Next, in Section 3, we discuss in more details the symmetries
of the theory, the analogue of Boulware transformation \cite{bou}
and the Ward identities corresponding to the
softly broken symmetry associated to torsion.
For pedagogical purposes, we simultaneously state
similar considerations for the vector field.
Section 4 is devoted to the calculation of the 1-loop divergences,
with many technicalities and the calculations for three 
simpler models are postponed to the Appendix A.
These calculations include the one for the massive vector
coupled to fermions. In order to perform calculations for the
cases of the massive vector and massive 
axial vector, we apply the generalized
Schwinger-DeWitt technique, developed in \cite{bavi}, which we
supplement by some technical tricks. The validity of the
calculational method is verified in two massless cases, for which
the result may be achieved through the Faddeev-Popov
method.
Section 5 contains further analysis of the 1-loop renormalization
and renormalization group equations. Section 6 is devoted to the
evaluation of the leading two-loop divergences. We apply, in this
section, the expansion of the loop integrals suggested in \cite{betor}.
Since the results of these two-loop calculations have great
importance for the qualitative output of our study, they are checked
in Appendix B by using the standard Feynman parameter method.
Finally, in Section 6, we draw our Conclusions.

\vskip 6mm
\noindent
{\large\bf 2. Dynamical torsion: review of previous results}
\vskip 2mm

In this section, we briefly present previous results.
We start off by the background notions for the
gravity with torsion
and quantum theory of matter fields in an external torsion field.
A pedagogical introduction may be found in \cite{book}.

In the space - time with independent metric and torsion, the affine
connection $\tilde{\Gamma}^\alpha_{\;\beta\gamma}$ is
non-symmetric, and the torsion tensor is defined as
$T^\alpha_{\;\beta\gamma} = {\tilde
{\Gamma}}^\alpha_{\;\beta\gamma} - {\tilde
{\Gamma}}^\alpha_{\;\gamma\beta} $. The covariant derivative,
$\tilde{\nabla}_\mu$, is based on the non-symmetric connection
$\tilde{\Gamma}^\alpha_{\;\beta\gamma}$, while the notation
$\,\nabla_\mu\,$ is kept for the Riemannian covariant derivative.
$\,$
From the metricity condition, $\,\tilde{\nabla}_\mu
g_{\alpha\beta} = 0\,$, the solution for the connection can be
easily found. It proves useful to divide the torsion field into
three irreducible components:
\beq
T_{\alpha\beta\mu} =
\frac{1}{3}\, \left( T_{\beta}\,g_{\alpha\mu} -
T_{\mu}\,g_{\alpha\beta} \right) - \frac{1}{6}\,
\varepsilon_{\alpha\beta\mu\nu} S^{\nu} + q_{\alpha\beta\mu}
\label{repra},
\eeq
where the last tensor satisfies the conditions
$\,q^\alpha_{\;\beta\alpha} = 0\;\;\;\;\;{\rm and}\;\;\;\;
\varepsilon^{\alpha\beta\mu\nu}q_{\alpha\beta\mu} =0$.

Let us now consider the interaction of torsion with matter fields.
The interaction between a Dirac field, $\psi$, and an external
gravitational field with torsion is described by the action:
\beq
S_{1/2}= i\,\int d^4x\,{\bar \psi}\, \left[ \,\ga^\al \,\left(\,
{\pa}_\al - i\,q \,V_\al+ i\,\eta\,\ga_5\,S_\al\,\right) - im
\,\right]\,\psi\,,
\label{gen}
\eeq
where $\eta$ is an arbitrary
parameter, which equals $\,1/8\,$ for the special case of minimal
coupling. For our purposes, it is useful to keep $\,\eta\,$
arbitrary. We have included the Abelian vector field, $\,V_\al\,$,
for the sake of further convenience.

The study of the renormalization of gauge models in an
external gravitational field with torsion
has been carried out in \cite{bush}. In the general case, the
theory includes gauge as well as scalar and fermion fields linked by corresponding
interactions (typical examples are the Standard Model or GUT's), the
non-minimal interaction with torsion proved necessary not only for
the spinor, but also for scalar fields. In the last case, the essential
(necessary for the renormalizability) interactions are described
by the action:
\beq
S_{sc} = \int d^4x\,\left\{\,\frac12\,g^{\mu\nu}\,\pa_\mu\ph\,\pa_\nu\ph
+\frac12\,m^2\,\ph^2 + \frac12\,\xi\,S_\mu S^\mu\,\ph^2
+ \frac12\,\xi_1\,R\,\ph^2 \,\right\}\,.
\label{scal}
\eeq
Here, $\,\xi, \xi_1\,$ are non-minimal parameters. On has to notice that
only the interactions with the axial vector, $\,S_\mu\,$, are
important in both cases (\ref{gen}) and (\ref{scal}). The
interaction of scalars with $q^\al_{\,\cdot\mu\nu}$ and both spinors
and scalars with $T_\mu$ may be introduced, but it is purely
non-minimal. In the sequel, we consider only the axial part, $S_\mu$, of
the torsion tensor. Also, since metric and torsion are
independent fields, and we are especially interested in the
torsion effects, in what follows we consider the flat metric only.

The problem of the action for the dynamical torsion field is crucially
important for all investigations of the gravity with torsion. In the
literature, one can meet
several different approaches for the construction of a torsion action
\cite{kibble,nevill,seznew,carfie,betor}. In particular, \cite{kibble}
started from the gauge principle for gravity
(similar ideas are very popular; see \cite{hehl1,hehl2} for a
comprehensive
review). In \cite{nevill,seznew}, the family of the high-derivative
metric-torsion actions, leading to theories without
unphysical massive spin-2 ghosts has been constructed. Therefore,
in these works, the guiding principle was the unitarity of the theory.
In the analysis of the physical significance of torsion, its
most important part is the axial component, $\,S_\mu\,$, for it 
is the component which couples to the fermions.
In \cite{carfie}, it was readily noticed that for the
study of the possible torsion effects at low energies,
only the second-derivative terms are indeed relevant.
Furthermore, in  \cite{carfie}, it was established that,
as usual for the vector field, the propagation of both the
transverse and the longitudinal parts of the axial vector $\,S_\mu\,$
unavoidably breaks unitarity (see, for example, \cite{vector}).
After that, in
\cite{carfie}, only the longitudinal part of $\,S_\mu\,$
has been considered, and torsion was thereby reduced to its
pseudoscalar piece. In \cite{betor}, the problem of consistency
had been formulated in a closed form, taking both aspects of
effective field theory into account.
The choice of the action for the dynamical torsion field should
be made in such a way that it leads to a unitary and
renormalizable effective quantum field theory.

Let us see how this principle can be applied to the fermion-torsion
interaction. Starting from (\ref{gen}), we may notice that this
action possesses two symmetries: the usual gauge one,
\beq
\psi' = \psi\,e^{i\al(x)}
,\,\,\,\,\,\,\,\,\,\,\,\,\,\,
{\bar {\psi}}' = {\bar {\psi}}\,e^{- i\al(x)}
,\,\,\,\,\,\,\,\,\,\,\,\,\,\,
V_\mu ' = V_\mu + {q}^{-1}\, \pa_\mu\al(x)
\label{trans1},
\eeq
and an additional symmetry which is softly broken by the spinor mass
\cite{betor,rytor}:
\beq
 \psi' = e^{i\ga_5\be(x)}\,\psi
,\,\,\,\,\,\,\,\,\,\,\,\,\,\,
{\bar {\psi}}' = {\bar {\psi}}\,e^{i\ga_5\be(x)}
,\,\,\,\,\,\,\,\,\,\,\,\,\,\,
S_\mu ' = S_\mu - {\eta}^{-1}\, \pa_\mu\be(x)\,.
\label{trans}
\eeq
In fact, the last symmetry is the key point allowing to set up
a unique form of the torsion action.
Even softly broken, this symmetry yields the appearance
of the transverse second-derivative counterterm $S_{\mu\nu}^2$
and (exactly because it is softly broken) the massive counterterm,
both coming, for instance, from a single fermion loop (see,
for example, our calculations in the next section).
Thus, if we wish to have a renormalizable effective field theory
for the torsion, these two terms must be included into
the action for a dynamical torsion. On the other hand, the
condition of unitarity forbids the third possible structure
\footnote{All other possible terms exhibit higher derivatives
or they are non-local.}, $\,(\pa_\mu S^\mu)^2\,$.

Therefore, the only chance to meet the conditions for the low-energy
renormalizability and unitarity is to choose the expression
\beq
S_{tor-fer} = \int d^4x\,\left\{\,
- \frac14\,S_{\mu\nu}S^{\mu\nu}
+ \frac12\,M^2\,S_\mu S^\mu
+ i\,{\bar \psi}\, [\,\ga^\al \,( {\pa}_\al
+ i\,\eta\,\ga_5\,S_\al) - im \,]\,\psi\,\right\}\,
\label{geral}
\eeq
as the torsion-fermion action.

Expression (\ref{geral}) shall be the main object of study
in the present paper. However, it is very instructive for us to see,
how the introduction of the scalar fields explicitly breaks
the above scheme. One can consult the second work of Ref. \cite{betor}
for a complete consideration. When one implements scalars, the
Yukawa interaction produces a rigid breaking of the
symmetry (\ref{trans}). This happens because the Yukawa coupling
is massless. As a result of this breaking, there are no restrictions
on the divergences coming from the diagrams including the Yukawa
vertex. As it was proved
in \cite{betor}, these diagrams really require the longitudinal
counterterm $\,(\pa_\al S^\al)^2\,$ at the two-loop level.
Of course, in order to have a renormalizable theory,
the term $\,(\pa_\mu S^\mu)^2$ might be introduced
into the torsion action but, as it was already mentioned,
this immediately breaks unitarity. Therefore, in the
torsion-fermion-scalar theory, there is a manifest conflict
between renormalizability and unitarity. This conflict resembles
the similar one which takes place in high-derivative
gravity \cite{stelle,book}. The difference is that, for gravity,
there is a massless mode which provides classical
effects through the propagation of graviton, while for the
torsion there are no massless modes, and if the lightest torsion
mode has a mass of the Planck order, then an independent torsion
field simply does not exist.

One can imagine several possibilities
to overcome the crisis between renormalizability and unitarity, as
described above. For instance, it is possible to search for an
extra symmetry providing the cancellation of the longitudinal
divergences. Another option is to restrict our considerations to
theories without fundamental scalars, such as Technicolour
or the Nambu-Jona-Lasinho models. In view of this, it becomes
especially important to investigate whether the fermion-torsion
system satisfies the consistency conditions.
In the present paper, we are going to make a complete
study of the conditions of  renormalizability and unitarity
for the fermion-torsion system without scalar fields.

In the next sections, we shall show that, unfortunately, despite
the breaking of the symmetry (\ref{trans}) is soft, the final
situation is very similar to the one with the scalar fields.
One may maintain the unitarity of the renormalized theory, but only at
the expenses of a very rigid limit on the torsion mass, which must be
much larger than the one of the fermions and much lighter than the
fundamental scale.

\vskip 6mm
\noindent
{\large\bf 3. Boulware's parametrization and the Ward identities}
\vskip 2mm

We need to perform an analogue of Boulware transformation
\cite{bou} in the fermion - axial vector system.
For pedagogical reasons, we first consider the usual
vector case, that is, repeat the transformation of \cite{bou}.
The action, in original variables, has the form:
\beq
S_{m-vec} = \int d^4x\,\left\{\,
- \frac14\,V_{\mu\nu}V^{\mu\nu}
+ \frac12\,M^2\,V_\mu V^\mu +
i \,{\bar \psi}\, [\,\ga^\al \,( {\pa}_\al
- i\,g\,V_\al) - im \,]\,\psi\,\right\},
\label{mvec}
\eeq
and, after the change of the field variables \cite{bou}:
\beq
\psi = \exp \left\{ \frac{ig}{M}\cdot\ph \right\}\cdot\chi
\,,\,\,\,\,\,\,\,\,\,\,\,\,\,\,\,\,
{\bar {\psi}} = {\bar {\chi}}\cdot
\exp \left\{ -\frac{ig}{M}\cdot\ph \right\}
\,,\,\,\,\,\,\,\,\,\,\,\,\,\,\,\,\,
V_\mu = V_{\mu}^{\bot} - \frac{1}{M}\,\pa_\mu\ph\,,
\label{bu1}
\eeq
the new scalar, $\ph$, is completely factored out:
\beq
S_{m-vec} = \int d^4x\,\left\{\,
-\frac14\,\left(V^{\bot}_{\mu\nu}\right)^2
+ \frac12\,M^2\,V^{\bot}_\mu V^{\bot\mu}
+ \frac12\,\pa^\mu\ph\pa_\mu\ph
+ i\,{\bar \chi}\,[\ga^\al \,( {\pa}_\al
+ i\,g\,V^{\bot}_\al) ]\,\chi \,\right\}.
\label{masvec}
\eeq

Let us now consider the fermion-torsion system given by the action
(\ref{geral}). The change of variables, similar to the
one in (\ref{bu1}), has the form:
\beq
\psi = \exp\left\{\frac{i\eta}{M}\,\gamma^5\,\ph\right\}\,\chi
\,,\,\,\,\,\,\,\,\,\,\,\,\,\,\,\,\,
{\bar {\psi}} =
{\bar {\chi}}\, \exp\left\{ \frac{i\eta}{M}\,\gamma^5\,\ph\right\}
\,,\,\,\,\,\,\,\,\,\,\,\,\,\,\,\,\,
S_\mu = S_{\mu}^{\bot} - \frac{1}{M}\,\pa_\mu\ph\, ,
\label{bu}
\eeq
where $\,S_{\mu}^{\bot}\,$ and $\,S^{\|}_\mu = \pa_\mu\ph\,$ are the
transverse and longitudinal parts of the axial vector respectively, the
latter being equivalent to the pseudoscalar $\ph$. One has to notice that,
contrary to (\ref{bu1}), but in full accordance with (\ref{trans}),
the signs of both the exponents in (\ref{bu}) are the same.
In terms of the new variables, the action becomes
$$
S_{tor-fer} = \int d^4x\,
\left\{\,
-\frac14\,S^{\bot}_{\mu\nu}S^{{\bot}\mu\nu}
+ \frac12\,M^2\,S^{\bot}_\mu S^{\bot\mu} +
\right.
$$
\beq
\left.
+ i\,{\bar \chi}\, [\,\ga^\al \,( {\pa}_\al
+ i\,\eta\,\ga_5\,S^{\bot}_\al\,)
- im\cdot e^{ \frac{2i\eta}{M}\,\gamma^5\,\ph}]\,\chi
+ \frac12\,\pa^\mu\ph\pa_\mu\ph \,\right\}\,,
\label{heral}
\eeq
where $\,S^{\bot}_{\mu\nu} =
\pa_\mu S^{\bot}_\nu - \pa_\mu S^{\bot}_\nu = S_{\mu\nu}\,$.
The last expression can be more easily analyzed by comparison
with a similar parametrization for the massive Abelian field
(\ref{bu1}).
Contrary to the last, for the torsion axial vector (\ref{heral}) the
scalar mode does not decouple, but rather couple with interactions
as follows:
\vskip 1mm

\noindent
i) Yukawa-type, resembling the problems with the ordinary scalar.

\noindent
ii) Exponential, which prevents the model from being power-counting
renormalizable.
\vskip 1mm

However, at first sight, there is a hope that the above features
would not be fatal for the theory. With respect to the point (i), one can
guess that the only result of the non-factorization, which
could be dangerous for the consistency of the effective quantum theory,
would be the propagation of the longitudinal mode of the torsion, and this
does not directly follow from the non-factorization of the scalar
degree of freedom in the classical action. On the other hand,
(ii) indicates the non-renormalizability, which might mean just the
appearance of the higher-derivative divergences, that do not matter
within the effective approach. Thus, a more detailed analysis
is necessary. In particular, the one-loop calculation in the theory
(\ref{geral}) may be helpful, and it will be done in the next section.

Let us consider, for a moment, the Ward-Takahashi identities 
for the two theories (\ref{mvec}) and (\ref{geral}.) In the 
case of the massive vector (\ref{mvec}), the identity for the 
effective action, $\,\Ga [V_\mu, \psi, {\bar \psi}]\,$, (here 
$\,V_\mu, \psi, {\bar \psi}\,$ are mean, or background fields) 
has the form 
\beq 
i\partial _{\mu}\frac{\de \Ga}{\de 
V_{\mu}}+ie\left(\bar{\psi} \frac{\de \Ga}{\de 
\bar{\psi}}-\frac{\de \Ga}{\de \psi}\psi\right) -iM^2\partial 
^{\mu}V_{\mu}=0\,.  \label{WTve} 
\eeq 
Applying the $\de / \de 
V_{\mu}$ operator, and setting $V_{\mu},\, \psi,\, 
 \bar{\psi}=0$, we get the identities for the inverse 
 propagator 
\beq 
\partial _{\mu}\frac{\de ^2 \Ga}{\de 
V_{\mu}(x)\de V_{\nu}(y)} = M^2\partial _{\nu}\de (x-y)\,.  
\label{WTve2}
\eeq
Now, applying $\de ^2 / \de\psi (y)\de\bar{\psi}(z)$, one obtains
\beq
\partial _{\mu}
\frac{\de^3 \Ga}{\de\psi (y)\de\bar{\psi}(z)\de V_{\mu}(x)}
=ie\left(\frac{\de ^2\Ga}{\de\psi (y)\de\bar{\psi}(z)}\de (x-y)-
\frac{\de ^2\Ga}{\de\psi (y)\de\bar{\psi}(z)}\de (x-z)\right)\,.
\label{WTve3}
\eeq
Similar relations take place for other vertices. The vector mass
completely decouples and shows up exclusively in the
propagator. Indeed, under these circumstances, it cannot affect the
divergences, except in some trivial way. The result is nothing but
the direct confirmation of the decoupling which is observed in
Boulware-like parametrization (\ref{masvec}).
\vskip 2mm

In our case of the massive axial vector field (\ref{geral}),
we have, by means of analogous procedures,
\beq
-\partial _{\mu}\frac{\de \Ga}{\de S_{\mu}}
-i\eta\left( \bar{\psi}\ga _5
\frac{\de \Ga}{\de \bar{\psi}}
-\frac{\de \Ga}{\de \psi}\ga _5\psi\right)
+2i\eta m\bar{\psi}\ga _5\psi + M^2\partial ^{\mu}S_{\mu}=0\,.
\label{WTax}
\eeq
Applying functional derivatives, the Ward-Takahashi identities
for the inverse propagator and the vertices take over the form:
\beq
\partial _{\mu}\frac{\de ^2 \Ga}{\de S_{\mu}(x)\de S_{\nu}(y)} =
M^2\partial _{\nu}\de (x-y)
\label{WTax2}
\eeq
and
\beq
& & \partial _{\mu} \frac{\de ^3 \Ga}
{\de\bar{\psi}(z)\de S_{\mu}(x)\de\psi (y)}
=-2i\eta m\ga _5\de (x-y)\de (x-z)+ \\ \nonumber
& + & i\eta\left(
-\frac{\de ^2\Ga}{\de\bar{\psi}(z)\de\psi (y)}\ga _5\de (x-y)
+\de (x-z)\ga _5\frac{\de ^2\Ga}{\de\psi (y)\de\bar{\psi}(z)}\right)
\label{WTax3}\,,
\eeq
and so on. The last expressions manifest the clear difference with respect
to the previous ones, (\ref{WTve}) -- (\ref{WTve3}). In the axial vector
case, the massive term affects the interaction vertices, and
one can expect that some non-invariant divergences may show up.

\vskip 6mm
\noindent
{\large\bf 4. One-loop calculation in the fermion-torsion case.}
\vskip 2mm

The purpose of this section is to derive the full set of 
1-loop counterterms
for the massive axial vector coupled to the Dirac spinor.
To get them, we are going to apply the background field method
together with the Schwinger-DeWitt expansion. However, since the use
of these methods for the system of interest
is quite non-trivial, and also for pedagogical
reasons, we perform also three auxiliary calculations: for the massless
vector coupled to a massive spinor (QED), for the
massless axial vector (this one coupled to massless
spinors) and for the massive vector, all using the same calculational
scheme as for the case of the massive axial vector. 
These additional
calculations are collected in the Appendix A. Here, we present the
details of calculation for the massive axial vector.

Let us start from the fermionic determinant, which was already
considered by many authors \cite{gold,buodsh,cogzer,betor}.
The contribution from the single fermion loop is given by
the expression
\beq
\Ga^{(1)}_{fermion} \,= \,- i \,Tr\ln\,{\hat H}\,,
\,\,\,\,\,\,\,\,\,
{\rm where} \,\,\,\,\,\,\,\,\, {\hat H} = \{i\ga^\mu D_\mu +m\}\,.
\label{fer-det}
\eeq
Here, $D_\mu = \pa_\mu + i \eta\ga^5 S_\mu$
is the covariant derivative.
It proves useful to introduce the conjugate derivative,
$D^*_\mu = \pa_\mu - i \eta\ga^5 S_\mu$. Then, one can write
$$
\Ga_{fermion} = - \frac{i}{2} \,Tr\ln\,{\hat H}\cdot{\hat H}^* =
 - \frac{i}{2} \,Tr\ln\,\{- \ga^\mu D_\mu\ga^\nu D_\nu - m^2\} =
$$
\beq
= -
\frac{i}{2} \,Tr\ln\,\{\,- (\ga^\mu\ga^\nu D^*_\mu D_\nu + m^2)\,\}\,.
\label{fer}
\eeq
After a simple algebra, one can cast two useful forms for the
operator between parenthesis: the non-covariant:
$$
-\,{\hat H}\cdot{\hat H}^* = \pa^2 + R^\mu\pa_\mu + \Pi\,,
$$
with
\beq
R^\mu =  2\eta\si ^{\mu\nu}\ga _5 S_{\nu}
\,,\,\,\,\,\,\,\,\,\,\,\,\,\,\,
\Pi = i\eta\ga _5(\partial _{\mu}S^{\mu})
+\frac{i}{2}\eta\ga ^{\mu}\ga ^{\nu}\ga _5 S_{\mu\nu}
+\eta^2 S_{\mu}S^{\mu} + m^2\, ;
\label{eta5}
\eeq
and covariant
\beq
-\,{\hat H}\cdot{\hat H}^* = D^2 + E^\mu D_\mu + F,
\label{eta55}
\eeq
\beq
{\rm with }\,\,\,\,\,\,\,\,\,\,\,\,\,\,
E^\mu = 2\eta\si ^{\mu\nu}\ga _5 S_{\nu} - 2i\eta\ga^5 S^\mu
\,,\,\,\,\,\,\,\,\,\,\,\,\,\,\,
F = m^2 + \frac{i}{2}\,\eta\ga^{\mu}\ga^{\nu}\ga^5 S_{\mu\nu}\,.
\label{cova}
\eeq
Both expressions are compatible with the use of the standard 
Schwinger-DeWitt technique (covariant calculation is much shorter),
which yields the well-known result
\beq
\Ga^{(1)}_{fermion,div} = \frac{\mu^{n-4}}{\varepsilon}
\int d^n x
\,\left\{ \frac{2\eta^2}{3}S_{\mu\nu}S^{\mu\nu}
- 8m^2\eta^2S^\mu S_\mu +2m^4 \right\}\,.
\label{fer-div}
\eeq
Here, $\,\varepsilon = (4\pi)^2\,(n-4)\,$ is the parameter of
dimensional regularization.

Now, we are in a position to start the complete calculation of
divergences. The use of the background field method supposes the
split (shift) of the field variables into a background and a
quantum part. However, in the case of the (axial)vector-fermion
system, the simple shift of the fields leads to an enormous volume of
calculations, even for a massive vector. Such a calculation becomes
extremely difficult for the axial massive vector (\ref{geral}).
That is why we have invented a simple trick combining the
background field method with the Boulware transformation
(\ref{bu}) for the quantum fields. As we shall see in a moment,
our method makes the calculations reasonably simpler.

Let us divide the fields into background $(S_\mu,\psi,{\bar \psi})$
and quantum $(t^{\bot}_\mu,\ph,\chi,{\bar \chi})$ parts, according to 
what follows:
$$
\psi \to \psi ^{'}=e^{i\frac{\eta}{M}\ga _5\ph}\cdot(\psi +\chi)\,,
$$
$$
\bar{\psi} \to \bar{\psi}^{'}=
(\bar{\psi}+\bar{\chi})\cdot e^{i\frac{\eta}{M}\ga _5\ph}\,,
$$
$$
S_{\mu} \to S_{\mu}^{'}
=S_{\mu}+t^{\perp}_{\mu}-\frac{1}{M}\partial _{\mu}\ph\,.
$$
The one-loop effective action depends on the quadratic (in quantum
fields) part of the total action:
$$
S^{(2)}=\frac{1}{2}\int d^4x \,\left\{ t^{\perp}_{\mu}
(\,\Box + M^2\, )t^{\perp\mu} + \ph \,( - \Box )\,\ph
+ t^{\perp}_{\mu}\,(\, -2\eta\bar{\psi}\ga ^{\mu}\ga _5\, )\,\chi
+ \ph \,( - \frac{4m\eta ^2}{M^2}\,\bar{\psi}\psi )\,\ph +
\right.
$$
\beq
\left.
+ \bar{\chi}(\, -2\eta\ga ^{\nu}\ga^5\psi\, )\,t^{\perp}_{\nu}
+ \bar{\chi}\,(\frac{4im\eta}{M}\,\ga^5\psi )\,\ph
+\ph\, (\frac{4i\eta m}{M}\,\bar{\psi}\ga^5 )\,\chi +
\bar{\chi}\,(2i\ga ^{\mu}D_{\mu}+2m\,)\chi \right\} \,.
\label{bili}
\eeq
Making the usual change of the fermionic variables
\footnote{One has to remember that the Jacobian of this
change of variables has been already taken into account
before, and its divergences were counted in (\ref{fer-div}).},
$\,\chi = -\frac{i}{2}(\ga ^{\mu}D_{\mu}+im)\ta\,$,
and substituting $\,\ph \to i\ph$, we arrive at the following
expression:
$$
S^{(2)} \,=\,\frac{1}{2}\,\int d^4x \,
\left(\matrix{t^{\perp}_{\mu} & \ph & {\bar \chi} \cr}\right)
\cdot \,{\hat {\bf H}}\,\cdot
\left(\matrix{t^{\perp}_{\nu} \cr \ph \cr \tau \cr}\right)\,,
$$
where the Hermitian bilinear form $\,{\hat {\bf H}}\,$
has the form
\beq
{\hat {\bf H}} = \left( \begin{array}{ccc}
\th^{\mu\nu}(\Box + M^2) & 0 & \th ^{\mu}\mbox{}_{\be}
(L^{\be\al}\partial _{\al} + M^{\be}) \\
0 & \Box + N & A^{\al}\partial _{\al} + B \\
P_{\be}\th ^{\be\nu} & Q & \hat{1}\Box + R^{\la}\partial _{\la}+ \Pi
\end{array}\right)\,,
\label{2h}
\eeq
$\th^{\mu}\mbox{}_{\nu}=\de^{\mu}\mbox{}_{\nu}-\pa^{\mu}
\frac{1}{\Box}\pa_{\nu}$ being
the projector on the transverse vector states.
The elements of the matrix operator (\ref{2h}) are defined
according to (\ref{bili}). They include the expressions
(\ref{eta5}) and also
$$
L^{\al\be}=-i\eta \bar{\psi}\ga _5\ga ^{\al}\ga ^{\be}
\,,\;\;\;\;\;\;\;\;\;\;\;\;\;\;
M^{\be} = \eta ^2\bar{\psi}\ga ^{\be}\ga ^{\al}S_{\al}+
\eta m\bar{\psi}\ga _5\ga ^{\be}\,,
$$
$$
A^{\al} = 2i\eta\frac{m}{M}\bar{\psi}\ga _5\ga ^{\al}
\,,\;\;\;\;\;\;\;\;\;\;\;\;\;\;
B = 2\eta ^2\frac{m}{M}\bar{\psi}\ga ^{\be}S_{\be}-
2\eta \frac{m^2}{M}\bar{\psi}\ga _5\,,
$$
\beq
N = 4\eta ^2 \frac{m}{M^2}\bar{\psi}\psi
\,,\;\;\;\;\;\;\;
P^{\be} = -2\eta\ga ^{\be}\ga _5\psi
\,,\;\;\;\;\;\;\;
Q = -4\eta\frac{m}{M}\ga _5 \psi\,.
\label{blocks}
\eeq

The operator ${\hat {\bf H}}$ given above might look like the
minimal second order operator ($\,\Box + 2h^\la\na_\la + \Pi$); 
but, in fact, it is not minimal because of the projectors
$\th^{\mu\nu}$ in the axial vector- axial vector
$\,t^{\perp}_{\mu}$ -- $t^{\perp}_{\nu}\,$ sector.
That is why one cannot directly
apply the standard Schwinger-Dewitt expansion to derive the
divergent contributions to the one-loop effective action,
and some more sophisticated technique is needed.

Let us perform the expansion in the transverse axial vector space,
and then apply the generalized Schwinger-Dewitt technique
developed by Barvinsky and Vilkovisky \cite{bavi}. To some extent,
the transformations which we are going to do are similar to
the ones which have been used
for the calculations in high-derivative gravity coupled
to matter \cite{bush-high} (see also \cite{book}).
Notice that, in the present case, these transformations enable one
to perform the calculations in the Abelian vector theory.
For the massless case, the results are indeed the same as
the ones derived with the use of the Faddeev-Popov method.

Since we are dealing with the mixed operator including the boson
and fermion sectors, the trace of all products should be
understood as a supertrace ($Str$), which implies a positive
sign for the bosonic sector and negative sign for the fermionic sector.
One can perform the following expansion:
\beq
\Ga^{(1)} = \frac{i}{2}\sTr ln {\bf H} =
\frac{i}{2}\sTr ln \left( \begin{array}{ccc}
\th\Box & 0 & 0 \\ 0 & \Box & 0 \\ 0 & 0 & \hat{1}\Box
\end{array} \right)
-\frac{i}{2}\sTr\left\{\sum^{\infty}_{n=1}
\frac{(-1)^n}{n}\,\,(\,{\hat \Pi}\,\frac{1}{\Box})^n\, \right\}\,,
\label{expans2}
\eeq
where the operator $\,{\hat \Pi}\,$ corresponds to (\ref{2h}):
\beq
{\hat \Pi}\,\frac{1}{\Box} = \left\{ \begin{array}{ccc}
\th ^{\mu\nu}M^2\frac{1}{\Box} & 0 &
\th ^{\mu}\mbox{}_{\be}(L^{\be\al}\partial _{\al}
+ M^{\be})\frac{1}{\Box} \\
0 & N\frac{1}{\Box} &
A^{\al}\partial _{\al}\frac{1}{\Box}+B\frac{1}{\Box} \\
P^{\be}\th _{\be}\mbox{}^{\nu}\frac{1}{\Box} & Q\frac{1}{\Box} &
(R^{\la}\partial _{\la}+ \Pi)\frac{1}{\Box}
\end{array} \right\}.
\eeq
We are going to use the universal traces of \cite{bavi}, and since we
are working in flat space-time, the only non-zero traces,
for any given $n$, are
\beq
\Tr (\,
\partial _{\mu_1}\,\, ... \,\, \partial _{\mu_{2n-4}}
\frac{1}{\Box ^n}\, )|_{div} = -\frac{2i}{\vp}\int d^4x \,\,
\,\,\frac{g^{(n-2)}_{\mu_1 \, ... \, \mu_{2n-4}}}{2^{n-2}(n-1)!}\,.
\label{bavi}
\eeq
Here, the standard notational conventions of \cite{bavi} are used:
$$
g^{(0)}=1\,,\;\;\;\;\;\;\;\;\;
g^{(2)}_{\mu\nu}=g_{\mu\nu} \,,\;\;\;\;\;\;\;\;\;
g^{(4)}_{\mu\nu\al\be} = g_{\mu\nu}g_{\al\be} + g_{\mu\al}g_{\nu\be} +
g_{\mu\be}g_{\nu\al}\,,\;\;\;\;\;\;\;\;\;  {\rm e.t.c.}\,.
$$
It is easy to see, by counting the number of derivative in
the terms of the series (\ref{expans2}), that the divergences appear
only for $n=2,3,4$ and that the ones coming from $n=4$ are
completely defined by the fermionic operator (\ref{fer-det}), which
we have already taken into account. Therefore, now we only need to
work with the terms with $n=2,3$.

Consider the $n=2$ term.
\beq
\left(\,\hat{\Pi}\frac{1}{\Box}\,\right)^2 =
\left( \begin{array}{ccc}
D_1^{(2)} & E_1^{(2)} & E_2^{(2)} \\
E_3^{(2)} & D_2^{(2)} & E_4^{(2)} \\
E_5^{(2)} & E_6^{(2)} & D_3^{(2)} \end{array} \right)\,,
\eeq
where
\beq
D_1^{(2)} =
\th ^{\mu\nu}M^4\frac{1}{\Box ^2} +
\th ^{\mu}\mbox{}_{\be}L^{\be\al}\partial _{\al}\frac{1}{\Box}
P^{\ga}\th _{\ga}\mbox{}^{\nu}\frac{1}{\Box}
+ \th ^{\mu}\mbox{}_{\be}M^{\be}\frac{1}{\Box}
P^{\ga}\th _{\ga}\mbox{}^{\nu}\frac{1}{\Box}\,,
\eeq
\beq
E_1^{(2)} =
\th ^{\mu}\mbox{}_{\be}L^{\be\al}\pa_{\al}\frac{1}{\Box}Q\frac{1}{\Box}+
\th ^{\mu}\mbox{}_{\be}M^{\be}\frac{1}{\Box}Q\frac{1}{\Box}\,,
\eeq
\beq
E_2^{(2)} & = &
\th ^{\mu\rho}M^2\frac{1}{\Box}
\th _{\rho\be}L^{\be\al}\partial _{\al}\frac{1}{\Box} +
\th ^{\mu\rho}M^2\frac{1}{\Box}
\th _{\rho\be}M^{\be}\frac{1}{\Box} +
\th ^{\mu}\mbox{}_{\be}L^{\be\al}\partial _{\al}\frac{1}{\Box}
R^{\la}\partial _{\la}\frac{1}{\Box} + \\ \nonumber
& + &
\th ^{\mu}\mbox{}_{\be}L^{\be\al}\partial _{\al}\frac{1}{\Box}
\Pi\frac{1}{\Box} +
\th ^{\mu}\mbox{}_{\be}M^{\be}\frac{1}{\Box}
R^{\la}\partial _{\la}\frac{1}{\Box} +
\th ^{\mu}\mbox{}_{\be}M^{\be}\frac{1}{\Box}\Pi\frac{1}{\Box}\,,
\eeq
\beq
E_3^{(2)} =
A^{\al}\partial _{\al}\frac{1}{\Box}
P^{\be}\th _{\be}\mbox{}^{\nu}\frac{1}{\Box} +
B\frac{1}{\Box}P^{\be}\th _{\be}\mbox{}^{\nu}\frac{1}{\Box}\,,
\eeq
\beq
D_2^{(2)} =
N\frac{1}{\Box}N\frac{1}{\Box} +
A^{\al}\partial _{\al}\frac{1}{\Box}Q\frac{1}{\Box} +
B\frac{1}{\Box}Q\frac{1}{\Box}\,,
\eeq
\beq
E_4^{(2)} & = &
N\frac{1}{\Box}A^{\al}\partial _{\al}\frac{1}{\Box} +
N\frac{1}{\Box}B\frac{1}{\Box} +
A^{\al}\partial _{\al}\frac{1}{\Box}R^{\la}\partial _{\la}\frac{1}{\Box} +
A^{\al}\partial _{\al}\frac{1}{\Box}\Pi\frac{1}{\Box} + \\ \nonumber
& + &
B\frac{1}{\Box}R^{\la}\partial _{\la}\frac{1}{\Box} +
B\frac{1}{\Box}\Pi\frac{1}{\Box}\,,
\eeq
\beq
E_5^{(2)} =
P^{\be}\th _{\be}\mbox{}^{\rho}\frac{1}{\Box}
\th _{\rho}\mbox{}^{\nu}M^2\frac{1}{\Box} +
R^{\la}\partial _{\la}\frac{1}{\Box}
P^{\be}\th _{\be}\mbox{}^{\nu}\frac{1}{\Box} +
\Pi\frac{1}{\Box}P^{\be}\th _{\be}\mbox{}^{\nu}\frac{1}{\Box} \,,
\eeq
\beq
E_6^{(2)} =
Q\frac{1}{\Box}N\frac{1}{\Box} +
R^{\la}\partial _{\la}\frac{1}{\Box}Q\frac{1}{\Box} +
\Pi\frac{1}{\Box}Q\frac{1}{\Box}\,,
\eeq
\beq
D_3^{(2)} & = &
P^{\ga}\th _{\ga}\mbox{}^{\rho}\frac{1}{\Box}
\th _{\rho\be}L^{\be\al}\partial _{\al}\frac{1}{\Box}
+P^{\ga}\th _{\ga}\mbox{}^{\rho}\frac{1}{\Box}
\th _{\rho\be}M^{\be}\frac{1}{\Box} +
Q\frac{1}{\Box}A^{\al}\partial _{\al}\frac{1}{\Box} + \\ \nonumber
 & + &
Q\frac{1}{\Box}B\frac{1}{\Box} +
\left\{R^{\la}\partial _{\la}\frac{1}{\Box} +
\Pi\frac{1}{\Box} \right\}^2\,.
\eeq
Here, we use $\,\,\Tr (AB) = \pm \Tr (BA)\,\,$ for the operators,
depending on their Grassmann parity. Disregarding the purely
fermionic contributions (which we already calculated), we obtain
for the divergent part:
\beq
-\frac{1}{2}\sTr \left( \hat{\Pi}\frac{1}{\Box}\right) ^2 & = &
-\frac{1}{2}\Tr \left\{ \th ^{\mu\nu}M^4\frac{1}{\Box ^2} +
2\th ^{\mu}\mbox{}_{\be}L^{\be\al}\partial _{\al}\frac{1}{\Box}
P^{\ga}\th _{\ga}\mbox{}^{\nu}\frac{1}{\Box} + \right. \\ \nonumber
& + &
\left. 2\th ^{\mu}\mbox{}_{\be}M^{\be}\frac{1}{\Box}
P^{\ga}\th _{\ga}\mbox{}^{\nu}\frac{1}{\Box} + N^2\frac{1}{\Box ^2} +
2A^{\al}\partial _{\al}\frac{1}{\Box}Q\frac{1}{\Box} +
2B\frac{1}{\Box}Q\frac{1}{\Box} \right\}\,.
\eeq
After some involved commutations (which we do not discuss
because they are in fact similar to the ones
described in \cite{bush-high,book}), we arrive at
$$
-\frac{1}{2}\sTr \left( \hat{\Pi}\frac{1}{\Box}\right) ^2  =
-\frac{1}{2}\Tr \left\{ \th ^{\mu\nu}M^4\frac{1}{\Box ^2} -
4L^{\ga\al}(\partial ^{\rho}P_{\ga})\partial _{\al}\partial _{\rho}
\frac{1}{\Box ^3} +
2L^{\ga\al}(\partial _{\al}P_{\ga})\frac{1}{\Box ^2} +
\right.
$$$$
\left.
+ 4L^{\be\al}(\partial ^{\rho}P^{\ga})
\partial _{\al}\partial _{\be}\partial _{\ga}\partial _{\rho}
\frac{1}{\Box ^4} -
2L^{\be\al}(\partial _{\al}P^{\ga})\partial _{\be}\partial _{\ga}
\frac{1}{\Box ^3} +
2M^{\ga}P_{\ga}\frac{1}{\Box ^2} -
2M^{\be}P^{\ga}\partial _{\be}\partial _{\ga}\frac{1}{\Box ^3} +
\right.
$$
\beq
\left.
+ N^2\frac{1}{\Box ^2} -
4A^{\al}(\partial ^{\rho}Q)\partial _{\al}\partial _{\rho}
\frac{1}{\Box ^3} +
2A^{\al}(\partial _{\al}Q)\frac{1}{\Box ^2} +
2BQ\frac{1}{\Box ^2}\right\}\,.
\eeq
Using (\ref{bavi}), we get
\beq
-\frac{1}{2}\sTr \left( \hat{\Pi}\frac{1}{\Box}\right) ^2|_{div} & = &
\frac{i}{\vp}\int d^4x \left\{ 3M^4 + \frac{2}{3}L^{\ga\al}
(\partial _{\al}P_{\ga})+
\frac{1}{6}L_{\al}\mbox{}^{\al}(\partial _{\ga}P^{\ga}) + \right.
\\ \nonumber
& + &  \left. \frac{1}{6}L^{\ga\al}(\partial _{\ga}P_{\al})+
\frac{3}{2}M^{\ga}P_{\ga} +
A^{\al}(\partial _{\al}Q) + 2BQ + N^2 \right\}\,.
\eeq
By direct substitution of the expressions $\,A^\al,..,\Pi\,$,
and after some algebra, we arrive at the partial result:
\beq
-\frac{1}{2}\sTr \left( \hat{\Pi}\frac{1}{\Box}\right) ^2|_{div} & = &
\frac{i}{\vp}\int d^4x \left\{ 3M^4 +
16\eta^3\frac{m^2}{M^2}\bar{\psi}\ga _5 S\sla\psi +
(16\eta^2\frac{m^3}{M^2}-12\eta^2m)\bar{\psi}\psi +\right. \nonumber \\
& - &
\left. 6\eta^3\bar{\psi}\ga _5 S\sla\psi +
8i\eta^2\frac{m^2}{M^2}\bar{\psi}\partial\sla\psi +
16\eta^4\frac{m^2}{M^4}(\bar{\psi}\psi)^2  \right\}\,.
\label{parcial1}
\eeq
Note that each term inside this integral has the dimension of
$[mass]^4$, despite the unusual form of the contribution of the
torsion mass $M$.

Consider the $n=3$ term. Again, omitting all the contributions
into the fermionic sector, we write only the relevant terms:
\beq
\frac{1}{3}\sTr \left( \hat{\Pi}\frac{1}{\Box}\right) ^3|_{div} =
\frac{1}{3}\Tr \left(
3\th _{\mu\rho}L_{\rho}\mbox{}^{\al}\partial _{\al}\frac{1}{\Box}
R^{\la}\partial _{\la}\frac{1}{\Box}
P^{\be}\th _{\be}\mbox{}^{\nu}\frac{1}{\Box} +
3A^{\al}\partial _{\al}\frac{1}{\Box}R^{\la}\partial _{\la}\frac{1}{\Box}
Q\frac{1}{\Box}\right)\,.
\eeq
After some algebra, and using the universal traces (\ref{bavi}),
we obtain
\beq
\frac{1}{3}\sTr \left( \hat{\Pi}\frac{1}{\Box}\right) ^3|_{div} & = &
-\frac{2i}{3\vp}\int d^4x \left\{ \frac{5}{8}
L^{\rho\al}R_{\al}P_{\rho} -
\frac{1}{8}L^{\al}\mbox{}_{\al}R_{\la}P^{\la} + \right. \\ \nonumber
& - &
\left. \frac{1}{8}L^{\rho\al}R_{\rho}P_{\al} +
\frac{3}{4}A_{\al}R^{\al}Q \right\}\,.
\eeq
The relevant terms in the partial result ($n=3$) are
\beq
\frac{1}{3}\sTr \left( \hat{\Pi}\frac{1}{\Box}\right) ^3|_{div} =
\frac{i}{\vp}\int d^4x \left\{
6\eta ^3\bar{\psi}\ga _5 S\sla\psi -
24\eta ^3\frac{m^2}{M^2}\bar{\psi}\ga _5 S\sla\psi \right\}\,.
\label{parcial2}
\eeq
Summing up the contributions to the one-loop divergences
of (\ref{expans2}), coming from (\ref{parcial1}) and (\ref{parcial2})
and (\ref{fer-div}), we obtain the complete expression for the 
one-loop divergences:
$$
\Ga ^{(1)}_{div} \,= \,- \,\frac{\mu^{n-4}}{\varepsilon}\,
\int d^n x \,\left\{ -\frac{2\eta^2}{3}\,S_{\mu\nu}S^{\mu\nu}
+ 8m^2\eta^2\,S^\mu S_\mu - 2m^4 + \frac{3}{2}\,M^4 +
\right.
$$
\beq
\left.
+ \left( 8\eta^2\frac{m^3}{M^2} -
6\eta^2m\right) \bar{\psi}\psi +
8\eta^4\frac{m^2}{M^4}(\bar{\psi}\psi)^2 +
4i\eta^2\frac{m^2}{M^2}\bar{\psi}\ga^\mu D^{*}_\mu \psi \right\}\,.
\label{result}
\eeq
It is interesting to notice that the above expression
(\ref{result}) is not
gauge invariant. It is not difficult to see that the non-invariant
terms come as a contribution of the scalar $\ph$. This indicates
that, unlike the massive (Abelian) vector field (see Appendix A),
for the massive
axial vector the violation of the symmetry (\ref{trans}) is
not soft. Therefore, we have confirmed our previous analysis
based on the Ward-Takahashi identities.
More detailed consideration of the renormalization
is presented in the next section.

\vskip 6mm
\noindent
{\large\bf 5. Renormalization and renormalization group}
\vskip 2mm

The expression (\ref{result}) for the 1-loop divergences in the
theory (\ref{geral}) has two non-invariant pieces. The first one comes from
the $\bar{\psi}\ga^\mu D^{*}_\mu \psi$ term, which is not invariant
with respect to (\ref{trans}). In fact, this divergence produces
just a slight change in the renormalization of the coupling
constant $\eta$, so that the softly broken symmetry (\ref{trans})
can be maintained at the quantum level. The second term is essentially
non-invariant $\,({\bar \psi}\psi)^2$-structure.
The renormalizability
of the theory requires the $\,({\bar \psi}\psi)^2\,$ term to be
introduced into the classical action, so that the corresponding
counterterm can be removed by means of the renormalization of the
corresponding parameter. Indeed, one can calculate again the
1-loop divergences taking this term into account. On the other
hand, this is not necessary, because there is no one-loop
diagram containing this $\,({\bar \psi}\psi)^2$-vertex
which could contribute to the dangerous longitudinal divergence.
\footnote{ Let us remind the danger of the $(\pa_\al S^\al)^2$-type 
counterterm, which spoils both the
renormalizability and the unitarity of the theory. }
Unfortunately, such a diagram exists at the two-loop level.
We postpone the analysis of the two-loop diagrams to the next
section, and consider now, in some details, the one-loop
renormalization.

The appearance of the new $\,({\bar \psi}\psi)^2$-vertex
shows that the fermion-torsion theory cannot be
consistent even as an effective quantum field theory, at least
without some additional restrictions being imposed. Let us try to
introduce some additional restrictions on the value of the torsion
mass, $M$. Suppose $\,m \ll M$. This means that the torsion mass is
much (let us say, some orders) larger than the mass of any fermion
interacting with torsion. Alternatively, one can suppose that
torsion interacts only with massless spinors (this case is free from
any problem at the quantum level, but the existence of the massless
spinors in the SM is nowadays problematic) or very light fermions and
decouples, by definition, from heavy fermions. Since our
simplified consideration does not distinguish heavy and light
quarks, leptons etc, we just accept $\,m \ll M\,$ for a moment.
Then, both types of non-invariant counterterms carry very
small coefficients, proportional to $\,(m / M)^2$. Suppose we
include the "dangerous" interaction $({\bar \psi}\cdot \psi)^2$
into the action, but with a very small coupling of the order
$\,\la\sim m^2 / M^4$. This relation will not be violated by the
renormalization group running of the coupling $\,\la\,$, and hence
the renormalizability is achieved with a very weak coupling
$\,({\bar \psi} \psi)^2$. As we shall see in the next section,
the two-loop contribution to the dangerous longitudinal
counterterm $(\pa_\al S^\al)^2$ contains the 
$\,({\bar \psi}\psi)^2$-vertex. 
Therefore, one finds it possible to preserve
renormalizability if the $(\pa_\al S^\al)^2$-term is included into
the action (\ref{geral}) with a coefficient 
$\,\,b\sim ({m} / {M})^4$.

Formally, if the $(\pa_\al S^\al)^2$-structure is present,
unitarity is broken, since the corresponding degree of freedom
is a ghost. This term, along with the canonical kinetic term
$S_{\mu\nu}S^{\mu\nu}$, will unavoidably plague the spectrum
with unphysical modes: either a tachyonic or a negative-norm
state (ghost) excitation will show up as a spin-1 or a scalar
excitation. However, unitarity is still
ensured in the spinor sector of the theory; it may break only
in the $torsion-torsion$ sector. Let us consider some low-energy
amplitude involving {\it in}-states of the propagating
transverse torsion. In order to generate {\it out}-states of the
longitudinal torsion, one has to consider the diagrams with
corresponding vertices. Such vertices are absent at tree-level, 
and the ones, which involve a non-invariant
$({\bar \psi}\cdot \psi)^2$-interaction show up, as we shall see
in the next section, at the
second loop only. Then, the longitudinal
{\it out}-state is suppressed by the coefficient $(m / M)^4$.
Therefore, in the low-energy amplitudes of the torsion
(axial vector $S_\mu$) scattering, the unitarity is maintained
with the precision $(m / M)^4$.

Consider the one-loop renormalization and the corresponding
renormalization group in some details. The relations between bare
and renormalized fields and the coupling $\eta$ follow from
(\ref{result}):
$$
S_\mu^{(0)} = \mu^{\frac{n-4}{2}}\,S_\mu\,
\left(\, 1 + \frac{1}{\ep}\cdot\frac{4\eta^2}{3}\,\right)
\,,\,\,\,\,\,\,\,\,\,\,\,\,\, \psi^{(0)} =
\mu^{\frac{n-4}{2}}\,\psi\, \left(\, 1 +
\frac{1}{\ep}\cdot\frac{2\eta^2m^2}{M^2}\,\right)\,,
$$
\beq
\eta^{(0)} = \mu^{\frac{4-n}{2}}\, \left( \,\eta -
\frac{1}{\ep}\cdot \frac{4\eta^3}{3}\cdot \left[1+
6\frac{m^2}{M^2}\right]\,\right)\,. \label{renor}
\eeq
Similar
relations for the parameter $\,{\tilde \la} =
\frac{M^4}{m^2}\,\la\,$ of the $\,\la (\bar \psi\,\psi)^2\,$-
interaction, have the form: 
\beq {\tilde \la}^{(0)} =
\mu^{4-n}\,\left[ {\tilde \la} + \frac{8\eta^4}{\ep} +
\frac{16{\tilde \la} \eta^2 m^2}{M^2\,\ep}+
\frac{20{\tilde \la}\eta ^2}{3\ep}\right]\,.
\label{reno}
\eeq
These relations lead to a renormalization group equation
for $\eta$, which contains a new term proportional to $(m / M)^2$:
\beq 
(4\pi)^2\,\frac{d\eta^2}{dt} = \frac{8}{3}\,[1+
6\frac{m^2}{M^2}]\,\eta^4\,,\,\,\,\,\,\,\,\, \,\,\,\,\,\, \eta(0) =
\eta_0\,. \label{rg} 
\eeq 
Indeed, for the case $m \ll M$ and in
the low-energy region, this equation reduces to the one presented
in \cite{betor} (that is identical to the similar equation of
QED). In any other case, the theory of torsion coupled to the
massive spinors is inconsistent, and equation (\ref{rg}) is 
meaningless.

One can also write down the renormalization group equation for the
parameter $\,{\tilde \la}\,$ defined above. Using (\ref{reno}), we
arrive at the following equation:
\beq
(4\pi)^2\,\frac{d {\tilde \la} }{dt} = 8\,\eta^4\,.
\label{psi4}
\eeq
This equation confirms the
lack of a too fast running for this parameter. Indeed, all the last
consideration is valid only under the assumption that $m \ll M$ 
and has very restricted sense.

\vskip 6mm
\noindent {\large\bf 6. Two-loop diagrams}
\vskip 2mm

Let us investigate the 2-loop diagrams contributing to the
propagator of the axial vector, $\,S_\mu$. The question we intend
to answer is whether there are longitudinal divergences
at the two-loop level. Therefore, it is reasonable to start
from the diagrams which can exhibit $\,1/\ep^2$-divergences
\footnote{In this paper, we adopt the dimensional regularization.
All necessary integrals may be found in \cite{leibr,hela}},
and only if none of them are found, we explore the  
$\,1/\ep\,$-pole, which is always more complicated to calculate.

The leading $\,1/\ep^2$-two-loop divergences of the
mass operator for the axial vector $\,S_\mu\,$ come from two
distinct types of diagrams: the ones with the
$\,({\bar \psi}\psi)^2\,$-vertex
and the ones without this vertex. As we shall ensure,
the most dangerous diagrams are those with 4-fermion interaction.
As we have seen in the last two sections, this kind of interaction
is a remarkable feature of the axial vector
theory, which is absent in a massive vector theory.
Now, we shall calculate divergent $\,1/\ep ^2\,$-contributions 
from two diagrams with the $\,({\bar \psi}\psi)^2\,$-vertex, 
using the expansion suggested in \cite{betor}; later on, in
Appendix B, this calculation will be checked using Feynman
parameters.

Consider first the diagram of Figure 1.
This graph can be expressed, after making some commutations
of the $\ga$-matrices, as
\beq
\Pi^1_{\mu\nu}  =
- \la\eta ^2 \; tr\;\left\{ I_{\nu} \cdot I_{\mu}
\right\},
\label{produ1}
\eeq
where $\,\la\sim \frac{m^2}{M^4}\,$ is the coupling of the
four-fermion vertex,
the trace is taken over the Dirac spinor space and
\beq
I_{\nu}(p) = \int\frac{d^dp}{(2\pi)^d}
\frac{p\sla -m}{p^2-m^2}\,\, \ga _{\nu}\,\,\ga _5\,\,
\frac{p\sla -q\sla -m}{(p-q)^2-m^2}\, .
\label{I}
\eeq
Following \cite{betor}, we can perform the expansion
\beq
\frac{1}{(p-q)^2-m^2}
=\frac{1}{p^2-m^2}\,\sum_{n=0}^{\infty}\,(-1)^n\,
\left( \frac{-2p\, .\, q+q^2}{p^2-m^2}\right)^n \label{expans}\,.
\eeq
Now, as far as we are working within an effective field theory
framework, it is possible to  omit the powers of $q$ higher than $2$.
These terms can give contributions to the divergences, but only
to the ones with higher derivatives, and they are, therefore,
out of our interest.

When performing the integrations, we trace just the divergent
parts, thus arriving (using the integrals from \cite{hela})
at the expressions:
\beq
I_{\nu}=\frac{i}{\ep}\,\,
\left\{ -\frac{1}{6}q^2\ga _{\nu}-2m^2\ga _{\nu}-
\frac{1}{6}\ga _{\al}\ga _{\nu}\ga _{\be}q^{\al}q^{\be}+mq_{\nu}\right\}
+\,\, ... ,
\label{resultI}
\eeq
where the dots stand for the finite and higher-derivative divergent
terms. Substituting this into (\ref{produ1}), we obtain the
leading divergence of the diagram of Fig. 1:
\beq
\Pi^{1,div}_{\mu\nu}= -\frac{\la\eta^2}{\ep ^2} \,\,
\left\{+16m^4\eta _{\mu\nu}+\frac{28}{3}m^2q_{\mu}q_{\nu}-
\frac{16}{3}m^2q^2\eta _{\mu\nu}\right\}
+\,\, ...
\label{resultPi}
\eeq
This result shows that the construction of the first diagram
contains an $\,1/\ep^2\,$-longitudinal counterterm.

\vskip 2mm

Consider the second two-loop diagram depicted in Fig. 2.
Its contribution to the polarization operator,
$\,\Pi^2_{\mu\nu}\,$, is written, after certain transformations,
in the following way:
\beq
\Pi^2_{\mu\nu}= - \la\eta^2 \;\,tr\,\left\{ I_{\nu\mu}\cdot J  \right\}\,,
\label{produ2}
\eeq
where
\beq
I_{\nu\mu}=\int\frac{d^dp}{(2\pi)^d}\,
\frac{p\sla -m}{p^2-m^2}\,\, \ga _{\nu}\,\,
\frac{p\sla -q\sla +m}{(p-q)^2-m^2}\,\, \ga _{\mu}\,\,
\frac{p\sla -m}{p^2-m^2}
\eeq
and
\beq
J=\int\frac{d^dp}{(2\pi)^d}\,
\frac{k\sla -m}{k^2-m^2}\,.
\eeq
It proves useful to introduce the following $\ga$-matrix definitions:
$$
A_{\al\nu\be\mu\rho}=
\ga _{\al}\ga _{\nu}\ga _{\be}\ga _{\mu}\ga _{\rho}\,,
$$$$
B_{\al\nu\mu\be}=
- q^{\rho}\,\ga_\al\ga_\nu\ga_\rho\ga_\mu\ga_\be
+ m\,(\ga_\al\ga_\nu\ga_\mu\ga_\be
-\ga_\nu\ga_\al\ga_\mu\ga_\be - \ga_\al\ga_\nu\ga_\be\ga_\mu)
$$$$
C_{\al\nu\mu} = m^2\,(\ga_\nu\ga_\al\ga_\mu
- \ga_\nu\ga_\mu\ga_\al - \ga_\nu\ga_\al\ga_\mu) +
m q^\be\,(\ga_\nu\ga_\be\ga_\mu\ga_\al
+ \ga_\al\ga_\nu\ga_\be\ga_\mu)\,.
$$
$$
D_{\nu\mu}=-m^2q^{\be}\,\ga_\nu\ga_\be\ga_\mu
+m^3\,\ga_\nu\ga_\mu\,.
$$
Then, the first integral can be written as
\beq
I_{\nu\mu}=\int\frac{d^dp}{(2\pi)^d}\,\,
\frac{A_{\al\nu\be\mu\rho}\cdot p^{\al}p^{\be}p^{\rho}
+B_{\al\nu\mu\be}\cdot p^{\al}p^{\be}
+C_{\al\nu\mu}p^{\al}+D_{\nu\mu}}
{(p^2-m^2)^2\left(\, (p-q)^2-m^2\,\right)}.
\label{I2}
\eeq
Using the expansion (\ref{expans}), and disregarding higher
powers of $q$, as well as odd powers of $p$ in the numerator
of the resulting integral, one obtains, after using standard 
results \cite{hela}:
\beq
I_{\nu\mu}=\frac{i}{\ep}\; \left\{\;
\frac{1}{4}B^{\al}\mbox{}_{\nu\mu\al}+\frac{1}{12}
(A^{\al}\mbox{}_{\nu\al\mu\rho}+A^{\al}\mbox{}_{\nu\rho\mu\al}
+A_{\rho\nu\al\mu}\mbox{}^{\al})q^{\rho}\;\right\}+\; ...
\eeq
which gives, after some algebra,
\beq
I_{\nu\mu} = \frac{i}{\ep}\; \left\{ \;
m\,\ga_\mu\ga_\nu + 3m \eta_{\mu\nu}
-\frac{2}{3}\ga_{\rho}q^{\rho}\eta_{\mu\nu}
+ \frac{1}{3}\ga _{\mu}q_{\nu}+
\frac{1}{3}\ga _{\nu}q_{\mu} \right\} +\; ...
\eeq
The divergent contribution to $J$ is
\beq
J=-\frac{i}{\ep}\; m^3 +\; ...
\eeq
Now, the calculation of (\ref{produ2}) is straightforward:
\beq
\Pi _{\mu\nu}= \frac{\la\eta^2}{\ep^2}\;
8m^4\eta _{\mu\nu} +\; ...
\label{result2}
\eeq
As we see, this diagram does not contribute to the kinetic
counterterm
(with accuracy of the higher-derivative terms), and hence the
cancellation of the contributions to the longitudinal
counterterm coming from $\,\Pi^1_{\mu\nu}\,$ do not take place.
This result is reproduced in the Appendix B, with the help of
the Feynman parameters.

One has to notice that other two-loop
diagrams do not include the 
$\,({\bar \psi}\psi)^2\,$-vertex. Thus, even if those
diagrams contribute to the longitudinal counterterm,
the cancellation with $\,\Pi^1_{\mu\nu}\,$ should require
some special fine-tuning between $\,\la\,$ and $\,\eta\,$.
In fact, one can prove, without explicit calculation, that
the remaining two-loop diagrams of Figs. 3 and 4
do not contribute to the longitudinal $\,1/\ep^2\,$-pole.
In order to see this, let us notice that the leading
(in our case  $\,1/\ep^2\,$) divergence may be obtained
by consequent substitution of the contributions from
the subdiagrams by their local divergent components.
Since the local counterterms produced by the
subdiagrams of the two-loop graphs depicted
in Figs. 3 and 4 are minus the one-loop expression
(\ref{result}), the corresponding divergent vertices
are $\,1/\ep\,$ factor classical vertices. Hence, in the
leading $\,1/\ep^2\,$-divergences of the diagrams of Figs 3 and 4,
one meets again the same expressions as
in (\ref{result}). The result of our consideration is, therefore,
the non-cancellation of the $\,1/\ep^2\,$-longitudinal
divergence (\ref{resultPi}). This means that the theory
(\ref{geral}), without additional restrictions
on the torsion mass, like $\,m \ll M\,$, is inconsistent at the
quantum level.

\vskip 6mm
\noindent
{\large\bf 10. Conclusions}
\vskip 2mm

We have investigated, in more details than in the previous
works \cite{betor},
the quantum field theory of the fermion-torsion
system. The torsion is presented by its purely antisymmetric
part, equivalent to the axial vector $S_{\mu}$.
It was shown that renormalizability and unitarity
may be achieved only in the case of massless spinors coupled
to massless torsion, without scalar fields. According to
recent data on the neutrino oscillations, all existing
fermions have a non-zero mass. Probably, this means that
they also interact with the Higgs scalar. Thus, it
is clear that torsion cannot be implemented in a Standard
Model scenario or, at least, into its versions which are available to
the date.

Alternatively, one has to input very severe restrictions
on the torsion mass, which has to be much greater than the
mass of the heaviest fermion (say, t-quark, with a mass of
175 $GeV$), and use an
effective quantum field theory approach, restricting considerations to
the low-energy amplitudes only. This approach implies the
existence of a fundamental theory which is valid at higher
energies. The effective theory may be used only at the energies
essentially smaller than the typical mass scale of the
fundamental theory. If the mass of torsion is comparable
to this fundamental scale, all the torsion degrees of freedom
may be described directly in the framework of the fundamental
theory.

Hence, in order to have propagating torsion,
one has to satisfy a double inequality:
\beq
m_{fermion} \ll M_{torsion} \ll M_{fundamental}\,.
\label{double}
\eeq
Usually, the fundamental scale is associated with the Planck mass,
$M_{Pl} \approx 10^{19}\,GeV$
\footnote{As a by product, our study shows that if the
real fundamental scale is just a few orders above $TeV$,
there is no room for an independent propagating torsion.
Thus, one cannot incorporate torsion into the resent
discussion of the $Tev$-gravity (see, for instance, \cite{TEV}).},
and therefore we still have a
huge gap on the energy spectrum, which is not completely
covered by the present theoretical consideration. Of course,
this gap cannot be closed by any experiment, because
the mass of torsion is too big. Even the restrictions coming
from the contact experiments \cite{betor} achieve only the
region $\,M < 3 \,Tev\,$, and that
is not enough to satisfy (\ref{double}) for all the fermions
of the Standard Model. It is clear that the existence of a
torsion-interacting
fermions with mass of many orders larger than $m_t$ (like the
ones which are expected in many GUT's) can close the gap
on the particle spectrum and "forbid" an independent torsion.

The situation with torsion is similar to the one with
quantum gravity. In both cases, there is a conflict between
renormalizability (which lacks, in case of gravity,
for the Einstein theory) and
unitarity which is violated in high-derivative models
\cite{stelle,highderi}.
In some sense, this analogy is natural, because both the metric
and torsion represent the internal aspects of the space-time
manifold rather than usual fields. Therefore, one of the
options is to give up the quantization of these two
fields and consider them only as a classical background.
If one does not accept this option, it is possible to
consider both the metric and torsion as effective low-energy
interactions resulting from a more fundamental theory like
string. Both the metric and torsion result from string, but the
crucial difference is that metric has massless degrees of
freedom while torsion appears to have a mass in all known
versions of string theory \cite{GSW}. It is interesting
that the study of an effective quantum field theory for the metric
does not meet major difficulties \cite{don,weinberg},
while the consistency of the theory requires a lower-bound
(\ref{double}) on the torsion mass. One can guess that this is
more than an accidental coincidence.

Indeed, it is possible, that some new symmetries will be discovered,
which make the consistent quantum theory of the propagating
torsion possible.
However, in the framework of the well-established results,
the most natural supposition is perhaps that the torsion does not
exist as an independent field, or that it is purely classical
field which should not be quantized.

On the other hand, our study does not close the possibility
of having composite torsion, which can appear, for instance,
as a vacuum condensate of the light (or maybe even massless)
spinor fields. This possibility deserves, from our point of view,
special investigation.

\vskip 5mm
\noindent
{\large\bf Acknowledgments.}
\vskip 2mm
One of the authors (I.L.Sh.) is grateful to M. Asorey for the
explanations concerning the effective approach to quantum field
theory, and to A. Belyaev for discussions on the
phenomenological aspects of torsion. I.L.Sh. and J.A.H.-N.
acknowledge kind
hospitality at the Departamento de F\'{\i}sica, Universidade Federal
de Juiz de Fora and support by CNPq (Brazil). The work of I.L.Sh.
was also
supported in part by Russian Foundation for Basic Research under the
project No. 99-02-16617. G.B.P. is grateful to CNPq for the grant.
The authors express their gratitude to 
A. B. Penna-Firme for the useful discussions.

\vskip 6mm
\noindent
{\large\sl Appendix A.}

\noindent
{\bf Calculation of divergences for massive
vector, massless vector and axial vector coupled to fermions}
\vskip 2mm

All the calculations below shall be performed on a flat
background. Consider first the theory for massive vectors
with the action (\ref{mvec})
$$
S=\int d^4x\,\left\{\,-\frac{1}{4}\,V_{\mu\nu}V^{\mu\nu}
+\frac{1}{2}\,M^2V_{\mu}V^{\mu} + i\bar{\psi}(\ga ^{\mu}D_{\mu}
-im)\psi\,\right\}\,,
\eqno(A1)
$$
where
$D_{\mu}= \partial _{\mu}-igV_{\mu} \;\;\;\;\;{\rm and}\;\;\;\;\;\;\;
V_{\mu\nu}=\partial _{\nu}V_{\mu}-\partial_{\mu}V_{\nu}$.
In the framework of the background field method, one performs
the shift
$$
\psi \to \psi ^{'} = e^{i\frac{g}{M}\,\ph}(\psi +\eta)\,,
$$
$$
\bar{\psi}\to
\bar{\psi}^{'}=(\bar{\psi}+\bar{\eta})e^{-i\frac{g}{M}\ph}\,,
$$
$$
V_{\mu} \to V_{\mu}^{'}=V_{\mu}+t^{\perp}_{\mu}
+\frac{1}{M}\partial _{\mu}\ph\,.
$$
As the scalar field does not couple to any other field, the
$\ph$-sector can be successfully factored out.
The quadratic (in the quantum fields $t^{\perp}_{\mu},\eta,{\bar \psi}$)
part of the action is, after the change of the variables
$\eta = -\frac{i}{2}(\ga ^{\mu}D_{\mu}+im)\ta $, written in the form
$$
S^{(2)} = \frac{1}{2}\int d^4x \left( \begin{array}{cc}
t^{\perp}_{\mu} & \bar{\eta} \end{array}\right) \, {\hat {\bf H}} \,
\left(\begin{array}{c} t^{\perp}_{\nu} \\ \ta \end{array}\right)\,,
\eqno(A2)
$$
where
$$
{\hat {\bf H}} = \left\{ \begin{array}{cc}
\th ^{\mu\nu}(\Box + M^2) &
\th ^{\mu}\mbox{}_{\be}(L^{\be\al}\partial _{\al} + N^{\be}) \\
h^{\be}\th _{\be}\mbox{}^{\nu} &  \Box + R^{\la}\partial _{\la}+ \Pi
\end{array} \right\}\,,
\eqno(A3)
$$
and
$$
L^{\be\al} = -ig\bar{\psi}\ga ^{\be}\ga ^{\al} \;,\;\;\;\;\;\;\;\;\;
N^{\be}=-g^2\bar{\psi}\ga^{\be}\ga^{\al}V_{\al}+mg\bar{\psi}\ga^{\be}\,,
$$
$$
h^{\be} = 2g\ga^{\be}\psi  \;,\;\;\;\;\;\;\;\;\;\;
R^{\la} = -2igV^{\la}\,,
$$
$$
\Pi = \frac{i}{2}\,g\ga ^{\mu}\ga ^{\nu} V_{\mu\nu}-
ig(\pa_{\mu}V^{\mu}) - g^2V^{\mu}V_{\mu}+m^2\,.
$$
The operator (A3) is simpler than the one in (\ref{2h}), because of the
decoupling of the scalar mode in the vector case.

Now, we can evaluate the one-loop divergences of the effective action
in the theory (\ref{mvec}). For this, we use the same expansion as
in (\ref{expans2}) but with
$$
\hat{\Pi}\frac{1}{\Box} = \left\{ \begin{array}{cc}
\th ^{\mu\nu}M^2\frac{1}{\Box} &
\th ^{\mu}\mbox{}_{\be}(L^{\be\al}\partial _{\al}
+ N^{\be})\frac{1}{\Box} \\
h^{\be}\th _{\be}\mbox{}^{\nu}\frac{1}{\Box} &
(R^{\la}\partial _{\la}+ \Pi)\frac{1}{\Box}
\end{array} \right\}\,,
\eqno(A4)
$$
and, looking for logarithmic divergences, restrict our
consideration to the terms with $n = 2,3,4$. Also we notice that,
as for the axial vector, $n=4$ contributions are coming from the
fermion loop and can be easily derived by standard means
\cite{book}. The $n=2,3$ terms can be worked out exactly
like for the axial vector case, and the partial results for the
divergences read
$$
-\frac{1}{2}\sTr(\hat{\Pi}\frac{1}{\Box})^2|_{{\rm div}} =
\frac{i}{\vp}\int d^4x\,\left\{\,3M^4
+ 6g^3\bar{\psi}V\sla\psi + 12g^2m\bar{\psi}\psi\,\right\}
\eqno(A5)
$$
and
$$
\frac{1}{3}\sTr(\hat{\Pi}\frac{1}{\Box})^3|_{{\rm div}} =
\frac{i}{\vp}\int d^4x\,\left\{\,-6g^3\bar{\psi}V\sla\psi\,\right\} \,.
\eqno(A6)
$$
Adding the standard contribution from the fermion loop, we
arrive at the complete expression for the divergences
$$
\Ga^{(1)}_{div} =
- \frac{1}{\varepsilon}\int d^4 x\,
\left\{\,\,\frac{3}{2}\,M^4 + 6g^2 m \bar{\psi}\psi
- \frac{2g^2}{3}V_{\mu\nu}V^{\mu\nu}-2m^4
\right\}\,.
\eqno(A7)
$$
\vskip 3mm

In the cases of the
 massless vector coupled to spinor field (QED),
and the massless axial vector coupled to massless spinor, the
1-loop calculation can be done in a standard manner with the help of the
Faddeev-Popov method. However, in order to check our calculational
method, we performed these calculations in the same way as for the
massive cases. The most of the intermediate calculations
can be easily restored using the massive cases, so we shall
give just a main results.
\vskip 3mm

For the fermion coupled to the massless vector (QED),
the calculation is very simple and the divergences
have the well-known form:
$$
\Ga ^{(1)}_{div} = \frac{1}{\vp}\,\int d^4x\,\left\{
-\,6e^2m\,\bar{\psi}\psi + \frac{2e^2}{3}\,F^2_{\mu\nu}\,\right\}\,.
\eqno(A8)
$$
\vskip 3mm
Consider, in some more details,
the theory of the massless axial vector coupled to the
massless Dirac spinor. The classical action is
$$
S=\int d^4x\,\left\{\,-\frac{1}{4}\,S_{\mu\nu}S^{\mu\nu}
+i\bar{\psi}\ga ^{\mu}D_{\mu}\psi\right\}\,,
\eqno(A9)
$$
where the covariant derivative is the same as for the massive case.
This action is completely invariant under the gauge transformation
(\ref{trans}). Performing the change of variables and
applying the background field method,
as described in Section 4, we can write the
bilinear form of the action in the form
$$
{\hat {\bf H}} = \left( \begin{array}{cc}
\th ^{\mu\nu}(\Box + M^2) & \th ^{\mu}\mbox{}_{\be}
(L^{\be\al}\partial _{\al} + M^{\be}) \\
P_{\be}\th ^{\be\nu} & \hat{1}\Box + R^{\la}\partial _{\la}+ \Pi
\end{array}\right)\,,
\eqno(A10)
$$
where
$$
L^{\al\be}=-i\eta \bar{\psi}\ga _5\ga ^{\al}\ga ^{\be}
\;\,,\;\;\;\;\;\; \;\;\;\;\;\;
M^{\be} = \eta ^2\bar{\psi}\ga ^{\be}\ga ^{\al}S_{\al}
\;\,,\;\;\;\;\;\; \;\;\;\;\;\;
P^{\be} = -2\eta\ga ^{\be}\ga _5\psi \label{operadores}\,.
\eqno(A11)
$$
\vskip 2mm
 The expansion for $\,\,\frac{i}{2}\,\Tr ln {\hat {\bf H}}\,$
and the remaining calculations will produce almost the same
intermediate formulas as for the fermion-massive vector
calculation. The reason is that the matrices ${\hat {\bf H}}$
have many identical structures, the only difference lying on
the equations (A11) above.

We have, after substituting (A11) into the previous general formulae,
noticed that two contributions cancel
$$
- \frac{1}{2}\,\sTr
\left( \hat{\Pi}\frac{1}{\Box}\right) ^2|_{div} =
\frac{i}{\vp}\int d^4x
\left\{ \,- 6\eta ^3\bar{\psi}\ga _5 S\sla\psi \,\right\}
\eqno(A12)
$$$$
\frac{1}{3}\sTr \left( \hat{\Pi}\frac{1}{\Box}\right) ^3|_{div} =
\frac{i}{\vp}\int d^4x \left\{
6\eta ^3\bar{\psi}\ga _5 S\sla\psi \right\}\,,
\eqno(A13)
$$
and the general expression for the divergences is completely
defined by the fermion loop:
$$
\Ga ^{(1)}_{div} = \frac{1}{\vp}\,\int d^4x\,
\frac{2\eta^2}{3}\,\,S^2_{\mu\nu}\,.
\eqno(A14)
$$
It is indeed gauge invariant. The same divergence follows
from the standard calculation using the Faddeev-Popov method.

\vskip 5mm
\noindent
{\large\sl Appendix B.}

\noindent
{\large \bf Two-loop calculation using Feynman parametrization}
\vskip 2mm

Here, we start from the expression (\ref{I}), performing in the
denominator the Feynman parametrization:
$$
\frac{1}{ab}=\int\limits_0^1
\, \frac{dx}{\left\{\, ax+(1-x)b\,\right\}^2}
\eqno(B1)
$$
Following the standard procedures in dimensional regularization, we have
to change the integration variable as
$\,p\to p^{'}=p-qx$. After that, one meets some known integrals and get
$$
I_{\nu}=\frac{\Ga (\ep)}{(4\pi)^2}\int _0^1dx\,
\left\{\,\De \ga_\nu + \De ^{-\ep}D_{\nu}\,\right\},
\eqno(B2)
$$
where
$$
\De =q^2x(1-x)-m^2 \,,
$$$$
D_{\nu}=
\ga_\al\ga_\nu\ga_\be\,q^\al q^\be x^2 +
xp^\al\,(-2m\,\eta_{\nu\al}+
2m\,\ga_\al\ga_\nu - \ga_\al\ga_\nu\ga_\be q^\be)
+  mq^\al\,\ga_\nu \ga_\al -m^2\ga _{\nu}\, .
\eqno(B3)
$$
By direct computation of the above integral,
one arrives exactly at the result found by
the previous method, eq. (\ref{resultI}).
Indeed, the polarization operator calculated by these two methods
turn out to be the same, eq. (\ref{resultPi}).

Now, we recalculate the diagram of the Fig. 2.
Starting from (\ref{I2}), one has to perform the Feynman
parametrization
$$
\frac{1}{ab^2}=\int\limits_0^1\,
\frac{2(1-x)\,dx}{\left( ax+(1-x)b\right) ^3},
\eqno(B4)
$$
and the usual variable shift $\bar{p}=p-xq$. After proper algebraic
manipulations, one arrives at the expression:
$$
I _{\nu\mu}=2\int _0^1dx\, (1-x)\int\frac{d^dp}{(2\pi)^d}\,\,
\frac{(A_{\al\nu\be\mu\rho}+A_{\al\nu\rho\mu\be}+A_{\rho\nu\be\mu\al})
xq^{\rho}+ B_{\al\nu\mu\be}}
{(p^2-\De )^3}\;p^{\al}p^{\be}.
\eqno(B5)
$$
Using known integrals in dimensional regularization, and
performing the integration over $x$, we obtain
$$
I _{\nu\mu} =
\frac{i}{12\ep}
\, (\ga _{\nu}\ga _{\mu}\ga _{\rho}q^{\rho}
+ \ga _{\mu}\ga _{\rho}\ga _{\nu}q^{\rho}
+  \ga _{\rho}\ga _{\nu}\ga _{\mu}q^{\rho})
+ \frac{1}{4\ep}\,(12m\,\eta_{\mu\nu} - 4m\ga_\mu\ga_\nu
+ 2q^{\rho}\,\ga_\mu\ga_\rho\ga_\nu ) +\, ... ;
\eqno(B6)
$$
and as we already have $J$, we arrive at a final result identical
to (\ref{result2}).

\vskip 12mm

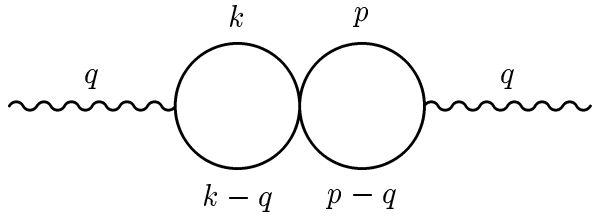
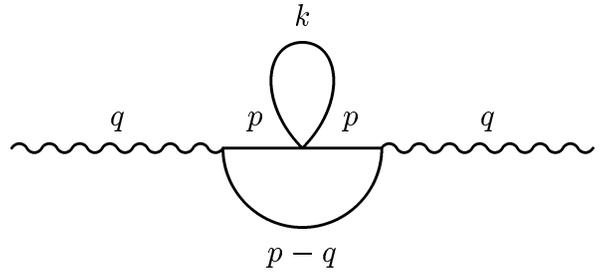
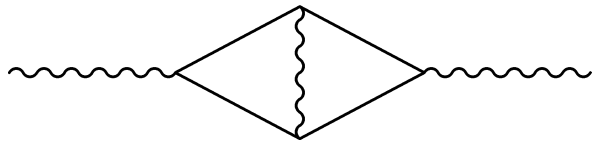
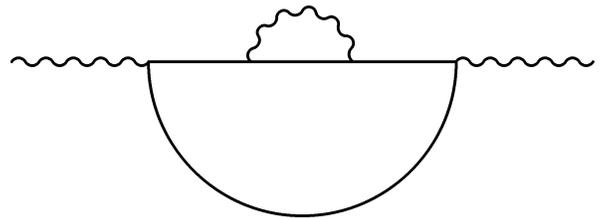
\end{document}